\documentclass[conference]{IEEEtran}
\IEEEoverridecommandlockouts

\usepackage[english]{babel}
\usepackage{cite}

\usepackage{bm}
\usepackage{bbm}
\usepackage{multirow}
\usepackage{adjustbox}

\usepackage{tikz}
\usetikzlibrary{arrows.meta}
\usetikzlibrary{backgrounds}
\usetikzlibrary{positioning, fit, mindmap, trees, calc,tikzmark,shapes}
\usetikzlibrary{shapes.arrows, fadings, automata,tikzmark,decorations.pathreplacing,patterns}
\usepackage{pgfplots}
\usepgfplotslibrary{groupplots}
\usepackage{xcolor}

\definecolor{riptide}{RGB}{141,211,199}
\definecolor{pale_prim}{RGB}{255,255,179}
\definecolor{lavender_gray}{RGB}{190,186,218}
\definecolor{salmon}{RGB}{242,131,107}
\definecolor{seagull}{RGB}{128,177,211}
\definecolor{rajah}{RGB}{253,180,98}
\definecolor{yellow_green}{RGB}{198,222,119}
\definecolor{classic_rose}{RGB}{252,205,229}
\definecolor{feijoa}{RGB}{178,223,138}

\definecolor{cruise}{RGB}{179,226,205}
\definecolor{apricot}{RGB}{253,205,172}
\definecolor{periwinkle}{RGB}{203,213,232}
\definecolor{snow_flurry}{RGB}{230,245,201}
\definecolor{buttermilk}{RGB}{255,242,174}

\definecolor{sundown}{RGB}{249, 180, 181}
\definecolor{spindle}{RGB}{179,205,227}
\definecolor{tea_green}{RGB}{204,235,197}
\definecolor{languid_lavender}{RGB}{222,203,228}
\definecolor{champagne}{RGB}{254,217,166}
\definecolor{cream}{RGB}{255,255,204}

\definecolor{monte_carlo}{RGB}{135,204,194}
\definecolor{melon}{RGB}{254,191,181}
\definecolor{granny_smith_apple}{RGB}{150,214,150}
\definecolor{watusi}{RGB}{254,221,207}
\definecolor{see_green}{RGB}{161,228,195}

\definecolor{moss_green}{RGB}{170,216,176}
\definecolor{opal}{RGB}{164,207,190}

\definecolor{pale_turquoise}{RGB}{172,240,242}
\definecolor{Madang}{RGB}{190,235,159}
\definecolor{pixie_green}{RGB}{183,214,170}
\definecolor{coral_andy}{RGB}{243,204,205}
\definecolor{manhattan}{RGB}{226,180,125}
\definecolor{quartz}{RGB}{219,223,238}
\definecolor{spring_sun}{RGB}{242,243,195}
\definecolor{dairy_cream}{RGB}{254,226,189}
\definecolor{surf_crest}{RGB}{205,230,208}
\definecolor{french_pass}{RGB}{195,232,246}
\definecolor{cosmos}{RGB}{248,209,210}
\definecolor{portafino}{RGB}{245,237,160}
\definecolor{sail}{RGB}{163,205,235}
\definecolor{hint_green}{RGB}{226,246,209}

\definecolor{jet_stream}{RGB}{188, 214, 210}

\definecolor{azalea}{RGB}{251, 196, 196}
\definecolor{wewak}{RGB}{244, 143, 150}
\definecolor{bittersweet}{RGB}{255,111,105}
\definecolor{sunset_orange}{RGB}{242,89,75}
\definecolor{light_coral}{RGB}{244, 127, 123}
\definecolor{carnation}{RGB}{245, 80, 86}
\definecolor{flamingo}{RGB}{237, 88, 85}
\definecolor{carmine_pink}{RGB}{231, 76, 60}
\definecolor{deep_carmine_pink}{RGB}{236, 50, 67}
\definecolor{fire_engine_red}{RGB}{210,44,41}
\definecolor{amaranth}{RGB}{234,46,73}
\definecolor{ku_crimson}{RGB}{243, 0, 25}
\definecolor{fire_engine_red}{RGB}{206, 37, 51}
\definecolor{copper_rust}{RGB}{155, 64, 74}

\definecolor{chilean_fire}{RGB}{215, 87, 44}

\definecolor{japanese_laurel}{RGB}{53, 116, 40}

\definecolor{turmeric}{RGB}{211, 178, 76}
\definecolor{saffron}{RGB}{249,193,62}
\definecolor{my_sin}{RGB}{255, 176, 59}
\definecolor{tree_poppy}{RGB}{246, 154, 27}
\definecolor{jaffa}{RGB}{240, 131, 58}
\definecolor{crusta}{RGB}{254, 127, 44}
\definecolor{tahiti_gold}{RGB}{223, 102, 36}
\definecolor{outrageous_orange}{RGB}{255, 100, 45}
\definecolor{safety_orange}{RGB}{254, 106, 0}

\definecolor{turquoise}{RGB}{41,217,194}
\definecolor{puerto_rico}{RGB}{94, 194, 166}
\definecolor{mountain_meadow}{RGB}{0, 163, 136}
\definecolor{free_speech_aquamarine}{RGB}{0, 156, 114}
\definecolor{java}{RGB}{2,190,196}

\definecolor{matisse}{RGB}{25, 104, 167}
\definecolor{shakespeare}{RGB}{85, 154, 193}
\definecolor{mona_lisa}{RGB}{246,152,134}
\definecolor{light_pink}{RGB}{255,182,193}

\definecolor{bgc}{RGB}{245,245,245}
\definecolor{tuatara}{RGB}{67, 67, 67}
\definecolor{aluminum}{RGB}{153,153,153}
\definecolor{silver}{RGB}{191,191,191}
\definecolor{platinum}{RGB}{228,228,228}
\definecolor{mercury}{RGB}{230,230,230}
\definecolor{gallery}{RGB}{240,240,240}
\definecolor{athens_gray}{RGB}{236, 240, 241}
\definecolor{ship_gray}{RGB}{77,77,77}

\definecolor{early_dawn}{RGB}{252,243,218}
\definecolor{egg_shell}{RGB}{238, 234, 215}
\definecolor{midnight}{RGB}{0, 29, 50}
\definecolor{sundown}{RGB}{249, 180, 181}
\definecolor{sun_shade}{RGB}{255, 144, 68}
\definecolor{sushi}{RGB}{117, 168, 47}
\definecolor{tomato}{RGB}{255, 97, 56}
\definecolor{ice_cold}{RGB}{169,232,220}

\definecolor{jelly_bean}{RGB}{45, 126, 150}
\definecolor{celestial_blue}{RGB}{52, 152, 219}
\definecolor{curious_blue}{RGB}{41, 128, 185}
\definecolor{french_blue}{RGB}{0, 112, 182}
\definecolor{matisse}{RGB}{25, 104, 167}

\definecolor{biscay}{RGB}{44, 62, 80}

\definecolor{cosmic_latte}{RGB}{222, 247, 229}
\definecolor{chinook}{RGB}{163, 232, 178}
\definecolor{padua}{RGB}{121, 189, 143}
\definecolor{ocean_green}{RGB}{79, 176, 112}
\definecolor{pastel_green}{RGB}{107, 227, 135}
\definecolor{chateau_green}{RGB}{69, 191, 85}
\definecolor{RoyalBlue}{RGB}{69, 191, 85}
\definecolor{pigment_green}{RGB}{0, 175, 79}
\definecolor{fern}{RGB}{101,197,117}
\definecolor{killarney}{RGB}{56, 113, 66}
\definecolor{viridian}{RGB}{70, 137, 102}

\usepackage{cite}
\usepackage{amsmath,amssymb,amsfonts}
\usepackage{algorithmic}
\usepackage{graphicx}
\usepackage{textcomp}
\usepackage{booktabs}
\usepackage{xcolor}
\usepackage{url}
\usepackage{enumitem}
\usepackage[numbers]{natbib}
\usepackage[para,online,flushleft]{threeparttable}
\def\BibTeX{{\rm B\kern-.05em{\sc i\kern-.025em b}\kern-.08em
    T\kern-.1667em\lower.7ex\hbox{E}\kern-.125emX}}
    
\renewcommand{\arraystretch}{1.25}
\newcommand\blfootnote[1]{%
  \begingroup
  \renewcommand\thefootnote{}\footnote{#1}%
  \addtocounter{footnote}{-1}%
  \endgroup
}

\begin{document}

\title{Explore User Neighborhood for Real-time E-commerce Recommendation}

\author{\IEEEauthorblockN{
Xu Xie$^{1}$~~~~~Fei Sun$^{2\,*}$~~~~~Xiaoyong Yang$^2$~~~~~~Zhao Yang$^2$~~~~~Jinyang Gao$^2$~~~~Wenwu Ou$^2$~~~~Bin Cui$^{1 \sharp\dagger}$}
\IEEEauthorblockA{\textit{$^1$School of EECS \& Key Laboratory of High Confidence Software Technologies (MOE), Peking University} \\
\textit{$\sharp$Center for Data Science, Peking University \& National Engineering Laboratory for Big Data Analysis and Applications} \\
\textit{$\dagger$Institute of Computational Social Science, Peiking University(Qingdao), China}
\textit{$^2$Alibaba Group}\\
\textit{$^1$\{xu.xie, bin.cui\}@pku.edu.cn}
\textit{$^2$\{ofey.sf, xiaoyong.yxy, haoyi.yz, jinyang.gjy, santong.oww\}@alibaba-inc.com}\\}
}

\maketitle

\begin{abstract}
Recommender systems play a vital role in modern online services, such as Amazon and Taobao.
Traditional personalized methods, which focus on user-item (UI) relations, have been widely applied in industrial settings, owing to their efficiency and effectiveness.
Despite their success, we argue that these approaches ignore local information hidden in similar users.
To tackle this problem, user-based methods exploit similar user relations to make recommendations in a local perspective.
Nevertheless, traditional user-based methods, like userKNN and matrix factorization, are intractable to be deployed in the real-time applications since such transductive models have to be recomputed or retrained with any new interaction.
To overcome this challenge, we propose a framework called self-complementary collaborative filtering~(SCCF) which can make recommendations with both global and local information in real time.
On the one hand, it utilizes UI relations and user neighborhood to capture both global and local information.
On the other hand, it can identify similar users for each user in real time by inferring user representations on the fly with an inductive model.
The proposed framework can be seamlessly incorporated into existing inductive UI approach and benefit from user neighborhood with little additional computation.
It is also the first attempt to apply user-based methods in real-time settings.
The effectiveness and efficiency of SCCF are demonstrated through extensive offline experiments on four public datasets, as well as a large scale online A/B test in Taobao.
\blfootnote{$^{*}$Corresponding author.}
\end{abstract}

\begin{IEEEkeywords}
Candidate generation, User neighborhood, Real time
\end{IEEEkeywords}

\section{Introduction}
Enormous amounts of products have induced great challenges to keep users satisfied in modern online services, such as e-commerce and social media.
Recommender systems play a crucial role in alleviating this information overload.
They usually suggest a small set of items from millions (even billions) of products, which are appealing to the specific user.
For large scale recommender systems, a common approach is a two-stage procedure, i.e., ``candidate generation'' step and ``ranking'' step. 
Candidate generation step selects a small subset of candidate items out of the entire item set for each user.
Then a ranking model predicts a score for each candidate item and select the top items based on the estimated scores.
This two-step procedure is widely adopted in the large scale industry recommender systems owing to its scalability and fast inference performance~\cite{slate,mind,sdm,dse1,dse2,scis1,scis2,fu,Miao}.
In this paper, we focus on the candidate generation stage~\cite{Covington:2016:DNN:2959100.2959190}, which is usually referred to as the \textit{top-$N$ recommendation}~\cite{Cremonesi:2010:PRA:1864708.1864721} in the academic area.

To cope with the candidate generation tasks, collaborative filtering~(CF) is one of the most pervasive methods.
It captures users' interests from their historical interactions~(e.g., clicks and purchases).
Through the information it focuses on, it can broadly fall into three categories: item-based approaches~\cite{Sarwar:2001:ICF:371920.372071, Linden:2003:ARI:642462.642471,Ning:2011:SSL:2117684.2118303}, user-item (UI) approaches~\cite{Covington:2016:DNN:2959100.2959190,Kabbur:2013:FFI:2487575.2487589,Koren:2008:FMN:1401890.1401944}, and user-based approaches~\cite{Herlocker:1999:AFP:312624.312682,Sedhain:2016:ELM:3015812.3015846,Sedhain:2016:PLM:3061053.3061158}.
Item-based approaches rely on item-item relations which are usually stable.
They compute the item similarity matrix in advance and then recommend items which are similar to what the users have interacted with.
While promising, these methods compute item similarities in a user-irrelevant manner, thus restricting their performance. 
To tackle this problem, UI methods takes more personalized information into account. 
They project users and items into a shared latent space and then recommend items which are close to the user in the space.

Despite the prevalence and effectiveness of UI approaches, we argue that they only capture the global structure of the item-item relations like item-based models~\cite{Christakopoulou:2018:LLS:3219819.3220112}.
There could exist pairs of items whose similarities tend to be small but display a strong correlation in a specific subset of users.
For example, consumers usually do not buy beer and diapers together, while new parents tend to purchase them together.
Therefore, though beer and diapers exhibit weak relations in a global aspect, they display strong relations in the new parents.
However, item-based methods and UI approaches project these pairs towards average position according to the whole users' interactions.
This drawback restricts their performances.

To address the above challenge, we attempt to model the local relations from a user-based aspect~\cite{Herlocker:1999:AFP:312624.312682,sarwar2000application, Sedhain:2016:ELM:3015812.3015846, Ebesu:2018:CMN:3209978.3209991,Koren:2008:FMN:1401890.1401944}.
For each user, we identify a set of users who have similar historical interactions, and then recommend items those similar users ever interacted with.
Traditional user-based methods, such as UserKNN~\cite{sarwar2000application}, identify similar users by computing user similarities directly from the original high-dimensional user interaction vectors.
It is not affordable with millions even billions of items.
To tackle this problem, some works factorize the user-item interaction matrix to reduce the dimension of user representations~\cite{ Sedhain:2016:ELM:3015812.3015846,Koren:2008:FMN:1401890.1401944}.
However, these methods can hardly be applied in real-world applications, since they usually fail to infer users' new preference in time.
\begin{figure}
  \centering
  \includegraphics[width=1.0\linewidth]{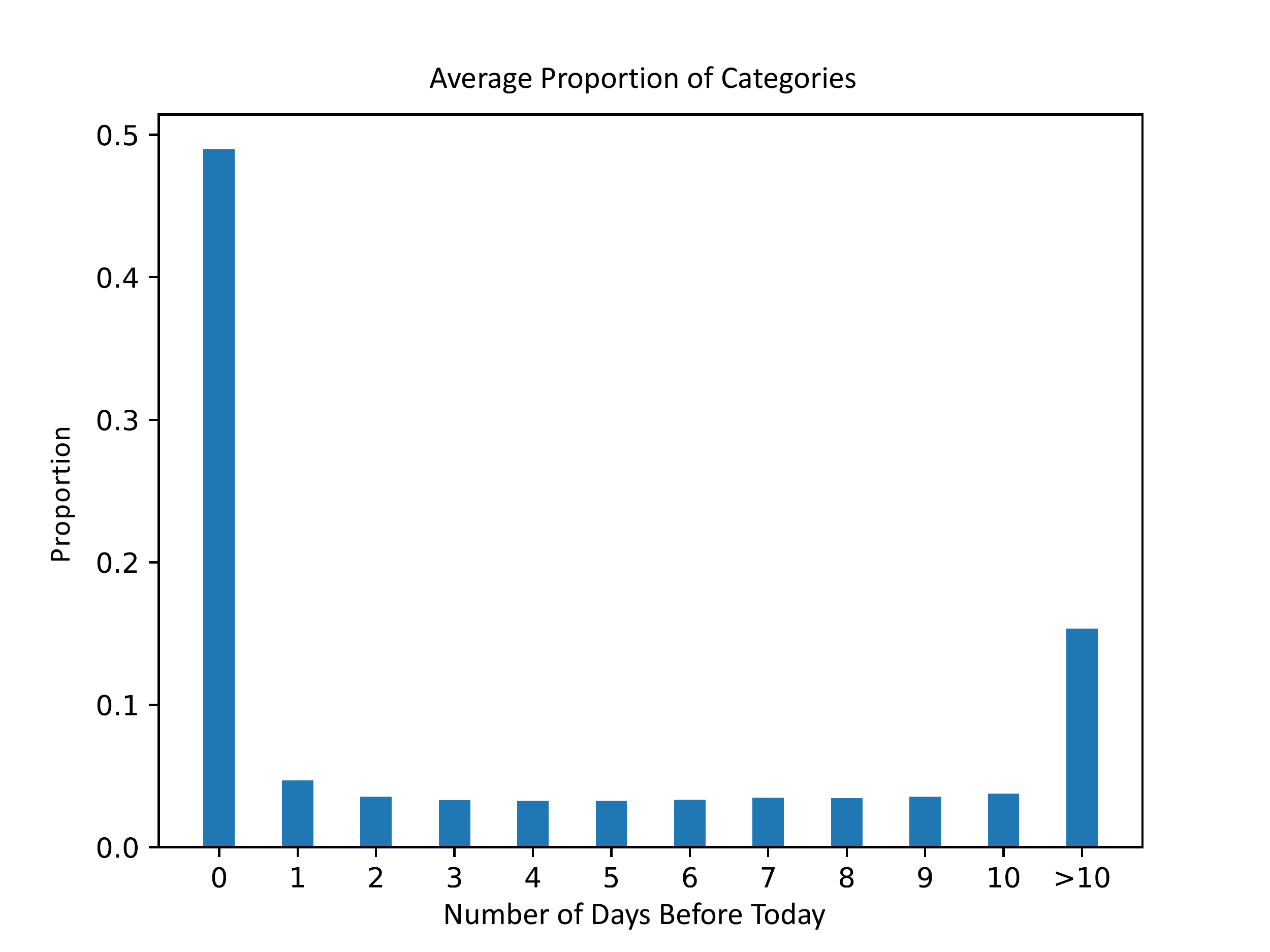}
  \caption{The average distribution of categories which a user clicks today and also clicked in the last two weeks. Each value at the day $x$ denotes how many categories she first clicked at $x$ days before today.}
  \label{Dynamic}
\end{figure}

To illustrate the necessity of the real-time issue, we analyze the changes of users’ interests on the Taobao platform.
We reflect this change by recording the proportion of categories that users click today and also clicked in the last two weeks. 
For each user, we record the number of categories that she clicks today.
And then, for each of these categories, we look for the day she first clicked this category in the last two weeks. 
The results are visualizing in Figure~\ref{Dynamic}. 
The value of $x$-axis denotes the number of days before today.
For example, the bar of $x=0$ demonstrates those categories are only clicked today and they were not clicked in the last two weeks.
The bar of $x = 4$ demonstrates those categories, which the user clicks today, are first clicked at 4 days before today. 
We observe that most of the categories, around 50\%, that users click today are new categories. 
Therefore, users' interests are sharply changing and we need to adapt to users’ dynamic interests in real time to make better recommendations.
However, when encountering with users' new interactions, these transductive models need to 
refactorize the new user-item matrix to infer new user representations, which is not affordable in the real-world services.

To tackle this challenge, we attempt to infer user representations in real time and then identify the similar users in real time.
With the recent advances of latent factor and deep learning techniques~\cite{Kabbur:2013:FFI:2487575.2487589, kang2018self,Covington:2016:DNN:2959100.2959190}, it is easy to update user vectors with new interactions in real time.
With the dynamically inferred user representations, 
users' neighborhood can be identified timely via efficient similarity search tools like \texttt{Faiss}~\cite{johnson2019billion}.
Specifically, we propose a general framework called \textbf{S}elf-\textbf{C}omplementary \textbf{C}ollaborative \textbf{F}iltering (\textbf{SCCF}) which captures both global and local relations in real-time settings.
There are three components in SCCF.
UI component derives user and item representations which encode complex global relations.
Then without any auxiliary parameters, user-based component identifies each user's appropriate neighborhood with the above representations, thus discovering the local patterns shared by her similar users. 
Finally, a neural network fuses the aforementioned results and derives the candidate list.
Hence, with the SCCF framework, we can complement the UI models with the local information hidden in user-user relations.
In addition, our SCCF framework can be regarded as a post-processing plugin to any inductive UI models whether shallow latent factor models or deep learning models, and whether sequential or non-sequential models.

Our primary contributions can be summarized as follows:
\begin{itemize}[leftmargin=2em]
\setlength\itemsep{0.3em}
    \item We revisit the user-based methods and reveal that it can enhance broadly used UI methods in a local aspect.
    \item Our framework addresses
    the real-time recommendation issues of traditional user-based approaches. To the best of our knowledge, we are the first to address this problem in the literature.
    \item We propose a framework called Self-Complementary Collaborative Filtering (SCCF), which can be seamlessly incorporated into inductive UI models and capture both global and local information.
    \item Extensive experiments on four public datasets demonstrate the effectiveness and efficiency of SCCF framework against state-of-the-art UI methods and user-based methods.
    \item We analyze the change of users' interests and also conduct an online A/B test on our e-commercial platform---Taobao.
    The results demonstrate the significance of capturing users' dynamic interests and verify the effectiveness of our SCCF framework.
\end{itemize}

The rest of the paper is organized as follows. 
We first list the relevant existing methods in section 2. 
Then, in section 3, we represent the details of SCCF framework and explain how to employ it to existing inductive UI models.
We provide the evaluation methodology, the datasets characteristics, and the experimental results on both public datasets and our online platform in secion 4 to support our claims.
Finally, we provide some concluding remarks in section 5.

\section{Related Work}
In this section, we review CF algorithms in the literature.
We classify them into three categories, including item-based models, user-item models, and user-based methods.

\subsection{Item-based Methods}

Item-based models~\cite{Sarwar:2001:ICF:371920.372071, Linden:2003:ARI:642462.642471, Deshpande:2004:ITN:963770.963776} estimate a user's preference on an item via measuring its similarity to the user's historical interacted items.
Since the item-item relations are usually stable, the item similarity table can be pre-built offline.
Then recommendations can be made in real time with just a series of look-up operations.
Owing to their efficiency and scalability, 
this type of methods is widely adopted in industry environments~\cite{Smith:2017:TDR:3101468.3101563,Davidson:2010:YVR:1864708.1864770,Gomez-Uribe:2015:NRS:2869770.2843948}.

There are two kinds of item-based methods including memory-based and model-based.
Memory-based methods calculate item similarities directly from the historical interactions.
They produce results fast with a sacrifice on recommendation quality.
To tackle this problem, model-based methods introduce learnable parameters to add the model capacity.
For example, \citet{Ning:2011:SSL:2117684.2118303} introduce an aggregation coefficient matrix learned by an linear regression to estimate item similarity~(SLIM).
\citet{GLSILM} extend SLIM by considering local item relations (GLSLIM).
Different from our SCCF which identifies each user's dynamic neighbors, GLSLIM clusters users into fixed groups, thus restricting its performance and failing to adapt to users' dynamic tastes.
Later, \citet{Barkan2016ITEM2VECNI} equip item-based methods with word2vec~\cite{Mikolov_word2vec:2013:DRW:2999792.2999959,Nedelec_content2vec:2017:SJR:3125486.3125489}.
It models items as latent vectors and computes item similarities by their inner-products.

\subsection{User-Item Representation Methods}
Item-item relations are usually user-independent, thus restricting the performance of item-based methods.
To tackle this problem, UI methods~\cite{Pan:2008:OCF:1510528.1511402, Pan:2009:MGW:1557019.1557094, Elkan:2008:LCO:1401890.1401920} focus on exploiting user-item relations and provide user-relevant recommendations.
Early works of UI methods leverage latent factor methods, such as Matrix Factorization~\cite{sarwar2000application, Salakhutdinov:2007:PMF:2981562.2981720,Paterek2007ImprovingRS}.
For example, \citet{Paterek2007ImprovingRS} projects users and items into a common latent space and then predicts the preference with the user-item inner-product (NSVD).
\citet{Kabbur:2013:FFI:2487575.2487589} improve NSVD by excluding diagonal entries to avoid the self-similarity problem (FISM).
SVD++~\cite{Koren:2008:FMN:1401890.1401944} captures users' interests more thoroughly by representing each user with user\_id and her/his interactions.

Recently, deep learning has been widely applied in recommender systems~\cite{Covington:2016:DNN:2959100.2959190, he2017neural}.
These works can be broadly classified into two categories: matching function based methods and representation based methods. 
Matching function based methods learn simple user/item representations and then utilize a complex matching model to capture user/item relations~\cite{he2017neural,he2018nais}.
Their high latency in the matching step is not affordable in the real-world scenario. 
Representation based methods focus on utilizing complex models to acquire meaningful user/item vectors, while their matching functions are usually very simple, like, inner-product, which is easy to be applied in real-world applications.
For example, \citet{Xue_deepMF:2017:DMF:3172077.3172336} extend MF with neural network to derive meaningful vectors~(DeepMF).
\citet{Chen:2017:ACF:3077136.3080797} adopt attention~\cite{DBLP:journals/corr/BahdanauCB14} mechanism to extract different weights on users' historical items and then obtain better user representations.
Meanwhile, some methods attempt to leverage sequence information in users' behaviors~\cite{kang2018self,Hidasi2015SessionbasedRW_gru4rec,Sun:cikm19:bert}.
For example, \citet{caser} propose a Convolutional
Sequence Model~(Caser) to learn sequential patterns using both
horizontal and vertical convolutional filters.
\citet{kang2018self} model long-term sequential semantics by self-attention mechanism~(SASRec), which achieves state-of-the-art performance.
In this paper, we focus on the Representation based manner.
We adopt the FISM and SASRec as our base model to evaluate the performance of our SCCF framework.

\subsection{User-based Methods}
Despite the success of UI methods, they just consider global item relations and neglect local correlations in the specific user neighborhood.
To tackle this problem, user-based methods~\cite{Herlocker:1999:AFP:312624.312682,sarwar2000application} seek to leverage user neighborhood to capture local patterns.
They first identify a set of similar users for each user and then recommend items that these similar users prefer.
Recently, many works attempt to enhance traditional user-based methods by introducing learnable parameters.
For example, \citet{Sedhain:2016:ELM:3015812.3015846} propose a variant of SLIM~(LRec) which learns a user-user similarity matrix and achieves better performance than traditional methods.
Many works~\cite{sarwar2000application,Sedhain:2016:PLM:3061053.3061158} also attempt to exploit latent factor models to estimate the user-user similarity through user representation embedding.
In addition, a few works attempt to incorporate user-based methods with deep learning approaches.
For example, \citet{Zhu_JCA:2019:ITR:3308558.3313678} leverage auto-encoders to jointly learn user-user and item-item correlations~(JCA).
However, JCA uses high dimensional sparse vectors as the input for users and items, which restrics it from being applied in the large scale industry scenario.
Later, \citet{Ebesu:2018:CMN:3209978.3209991} equip collaborative filtering with Memory Network (CMN)~\cite{DBLP_mem:journals/corr/WestonCB14}, which also takes user neighborhood into consideration. 

Despite the success of the above user-based methods, they fail to be applied in a real-time scenario where user-user relations are usually not stable~\cite{Smith:2017:TDR:3101468.3101563}.
Users' interests usually change sharply overtime.
Therefore, the similar users are also dynamic for each user.
To adapt to users' dynamic interests with their new interactions and identify new user-user relations, these methods need to be retrained since most of the above user-based methods are transductive models. 
However, retraining a model in a streaming manner is usually infeasible in real-world applications, thus restricting them from real-time candidate generation tasks.
Though CMN is an inductive model, it regards someone as the neighbor if she/he has any connection with the specific user's interacted items, which is also not affordable in real-world applications.
Different from the above methods, our SCCF addresses the real-time problems existing in the user-based literature by leveraging user neighborhood through the inductive UI model's representations.

\section{Self-Complementary Collaborative Filtering}

\begin{figure*}
  \centering
  \includegraphics[width=6in]{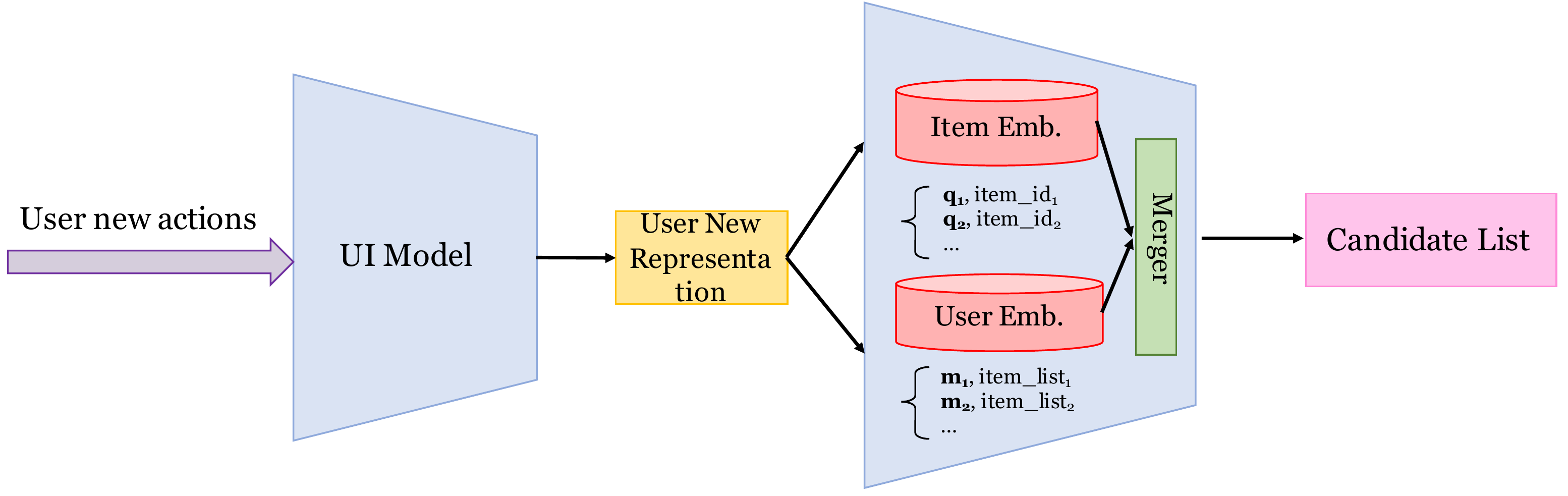}
  \caption{The pipeline of our proposed SCCF framework. When a user interacts with new items, the UI model of our SCCF first infers her/his new representation on the fly. Then our SCCF suggests two candidate lists in both global and local aspects. Finally the integrating component fuses together.}
  \label{framework of SCCF}
\end{figure*}

In this section, we propose our framework SCCF as illustrated in Figure~\ref{framework of SCCF}. 
In general, it consists of three components: a UI model, a user-based component, and an integrating component.
The UI model aims to obtain global user/item representations from users' historical behaviors.
The user-based component then discovers the local patterns through the above representations.
Finally, the integrating component fuses the above patterns and derives the candidate list. 
We first introduce notations used in the following parts and propose the research problem. 
Then we introduce SCCF in detail.

\subsection{Notations}
In this paper, all column vectors are represented by bold italic lower case letters (e.g., $\bm{p}$, $\bm{q}$).
All matrices are represented by bold upper case letters (e.g., $\bm{R}$).
The $i$-th row of a matrix $\bm{R}$ is represented by $\bm{R}_i^{\top}$.
We use calligraphic letters to denote sets (e.g., $\mathcal{U}$, $\mathcal{I}$). 

$\mathcal{U}$ and $\mathcal{I}$ denote the 
user/item set respectively, with $|\mathcal{U}| = n$ and $|\mathcal{I}| = m$. 
Our framework can be applied to non-sequential models and sequential models.
For non-sequential models, $\bm{R} \in \{0,1\}^{m \times n}$ is used to represent the user-item implicit feedback matrix.
$\mathcal{R}_u^+$ denotes the set of items that user $u$ has interacted with  and $\mathcal{R}_u^-$ is defined as $\mathcal{I}-\mathcal{R}_u^+$.
For sequential models, $\mathcal{S}_u=[i_1^{(u)},i_2^{(u)},\dots,i^{(u)}_{|\mathcal{S}^u|}]$ represents the interaction sequence in chronological order for  the user $u$.
And $i_t^{(u)}$
denotes the item which user $u$ interacts at the time step $t$.

To handle the candidate generation task, we aims to predict the most possible $N$ items which the user $u$ will click/purchase at the time step $|\mathcal{S}_u| + 1$.

\subsection{UI model}
\subsubsection{Infer User Representations}
Since our SCCF has no special requirements for the inductive UI methods,
we take two typical models as cases to introduce our framework, latent factor model FISM~\cite{Kabbur:2013:FFI:2487575.2487589} and deep sequential recommendation model SASRec~\cite{kang2018self}.

\textbf{FISM}.
Different from traditional matrix factorization methods, FISM just leverages users' history interactions to predict their preference without user identifier information.
It generates the representation for each user by aggregating the vectors of items which the user have interacted with.
Then, it predicts the recommendation score for a user on an unrated item by calculating the inner-product of their vectors.
For the input layer, one-hot encoding is applied to represent the ID feature of $\mathcal{R}_u^+$. 
And then, we adopt an embedding mechanism to map the sparse one-hot vector to an informative and lower-dimensional vector. 
In this way, for each item $i \in \mathcal{R}_u^+ $, we project it to an embedding vector $\bm{p}_i \in \mathbb{R}^d$. 
Therefore, the output of the input layer is a set of vectors $\mathcal{P}_u = \{\bm{p}_i|i\in\mathcal{R}_u^+\}$.

Then FISM copes with the vector set $\mathcal{P}_u$ as follows:
\begin{equation}
    \bm{m}_u=\frac{1}{(|\mathcal{R}_u^+|)^\alpha}\sum_{j\in\mathcal{R}_u^+}\bm{p}_j,
\end{equation}
where $\alpha$ is the normalization hyper-parameter that controls the degree of agreement between the items in $\mathcal{P}_u$\cite{Kabbur:2013:FFI:2487575.2487589}. 
If $\alpha$ is set to 1, it becomes the average pooling.
In contrast, with the value of 0, it turns to sum pooling . 

\tikzset{
  emb/.style = {draw, rectangle, fill=flamingo!62, minimum width=3em, minimum height=1.5em, inner sep=0, outer sep=0},
  transformer/.style = {draw, rectangle, fill=Madang, minimum width=3em, minimum height=1.5em, font=\scriptsize},
  norm/.style = {draw, rectangle, fill=spring_sun, minimum width=4.5em, minimum height=1em, inner sep=0, outer sep=0},
  dropout/.style = {draw, rectangle, fill=hint_green, minimum width=4.5em, minimum height=1em, inner sep=0, outer sep=0},
  mh/.style = {draw, rectangle, fill=dairy_cream, minimum width=4.5em, minimum height=1.5em},
  ff/.style = {draw, rectangle, fill=french_pass, minimum width=4.5em, minimum height=1.5em},
  posemb/.style = {draw, rectangle, fill=watusi, minimum width=3em, minimum height=1.3em, inner sep=0, outer sep=0},
  proj/.style = {draw, rectangle, fill=Madang, minimum width=3em, minimum height=1.5em, inner sep=0, outer sep=0},
  FARROW/.style={arrows={-{Latex[length=1mm, width=0.8mm]}}}
}

\begin{figure}
\centering
\resizebox{0.46\textwidth}{!}{
\begin{tikzpicture}

\node[] (e1) at (0, 0) {};
\node[right of=e1, node distance=1.2cm] (e2) {};
\node[right of=e2, node distance=1.2cm] (e3) {};
\node[right of=e3, node distance=2cm] (e4) {};

\node[emb, above of=e1, node distance=0.6cm, minimum height=1.2em] (sa1) {$\bm{p}_{i^{(u)}_1}$};
\node[emb, above of=e3, node distance=0.6cm, minimum height=1.2em] (sa3) {$\bm{p}_{i^{(u)}_{t-1}}$};
\node[emb, above of=e4, node distance=0.6cm, minimum height=1.2em] (sa4) {$\bm{p}_{i^{(u)}_{t}}$};
\foreach \x in {1,3,4}
{
\node[transformer, above of=sa\x, node distance=1cm, minimum height=1.2em, inner sep=0, outer sep=0] (st\x) {Trm};
\node[transformer, above of=st\x, node distance=1cm, minimum height=1.2em, inner sep=0, outer sep=0] (stt\x) {Trm};
}

\node[right of=sa1, node distance=1.2cm, minimum height=1.2em] (sa2) {$\ldots$};

\node[posemb, below of=sa1, node distance=0.8cm] (pe1) {$\bm{e}_1$};
\node[below of=sa2, node distance=0.8cm] (pe2) {$\cdots$};
\node[posemb, below of=sa3, node distance=0.8cm] (pe3) {$\bm{e}_{t-1}$};
\node[posemb, below of=sa4, node distance=0.8cm] (pe4) {$\bm{e}_{t}$};

\node[below of=pe1, node distance=0.7cm] (si1) {$i_1$};
\node[below of=pe3, node distance=0.7cm] (si3) {$i_{t-1}$};
\node[below of=pe4, node distance=0.7cm] (si4) {$i_{t}$};

\draw[FARROW] (si1.east) -- ++(0.45, 0) |-  (sa1.east) ;
\foreach \x in {1,3,4}
{
\node[below of=sa\x, node distance=0.4cm] (add\x) {\large\textbf{+}};
\draw[FARROW] (si\x.east) -- ++(0.45, 0) |-  (sa\x.east) ;
\draw[FARROW] (si\x) -- (pe\x) ;
}

\node[below of=pe2, node distance=0.7cm, minimum height=1.2em] (si2) {$\ldots$};
\node[above of=sa2, node distance=0.8cm, minimum height=1.2em] (st2) {$\ldots$};
\node[above of=st2, node distance=0.8cm, minimum height=1.2em] (stt2) {$\ldots$};
\node[above of=stt1, node distance=0.7cm] (so1) {$i_2$};
\node[above of=stt3, node distance=0.7cm] (so3) {$i_{t}$};
\node[above of=stt4, node distance=0.7cm] (so4) {$i_{t+1}$};

\foreach \x in {1,3,4}
{
\draw[FARROW] (sa\x.north) -> (st\x.south) ;
\draw[FARROW] (st\x.north) -> (stt\x.south) ;
\draw[FARROW] (stt\x) -> ($(so\x.south)-(0,-0.08)$) ;
}

\foreach \x/\y in {1/3, 1/4, 3/4}
{
\draw[FARROW] (sa\x.north) -> (st\y.south) ;
\draw[FARROW] (st\x.north) -> (stt\y.south) ;
}

\draw[thick,dotted]  ($(si1.south west)+(-0.4, 0)$) rectangle ($(so4.north east)+(0.4, 0)$);
\node[below of=si2, node distance=1.1cm, xshift=1cm] (sas) {(b) SASRec model architecture.};

\node[mh, left of=e1, node distance=2.7cm, align=center, font=\scriptsize, yshift=-0.3cm] (trm1) {Multi-Head\\ Attention};
\node[dropout, above of=trm1, node distance=0.75cm, align=center, font=\scriptsize] (do1) {Dropout};
\node[norm, above of=do1, node distance=0.6cm, align=center, font=\scriptsize] (n1) {Add \& Norm};

\draw[FARROW] (trm1) edge (do1);
\draw[FARROW] (do1) edge (n1);

\node[ff, above of=n1, node distance=1cm, align=center, font=\scriptsize] (ff1) {Position-wise\\Feed-Forward};
\node[dropout, above of=ff1, node distance=0.75cm, align=center, font=\scriptsize] (do2) {Dropout};
\node[norm, above of=do2, node distance=0.6cm, align=center, font=\scriptsize] (n2) {Add \& Norm};

\draw[FARROW] (n1) edge (ff1);
\draw[FARROW] (ff1) edge (do2);
\draw[FARROW] (do2) edge (n2);

\node[below of=trm1, node distance=1.2cm, align=center, font=\small] (input) {input};
\node[above of=n2, node distance=0.8cm, align=center, font=\scriptsize] (out) {};
\node[right of=out, node distance=0.7cm, align=center, font=\scriptsize, yshift=-0.35cm] (aa) {Trm};

\draw[FARROW] (n2) edge (out);
\draw[FARROW] (input) edge (trm1);

\draw[FARROW] ($(input.north)-(0,-0.3)$) -- ++(-1, 0) |- (n1.west) ;
\draw[FARROW] ($(n1.north)-(0,-0.3)$) -- ++(-1, 0) |- (n2.west) ;

\draw[thick,dotted, fill opacity=0.1, fill=bgc]  ($(input.south west)+(-0.8, 0.6)$) rectangle ($(out.north east)+(0.8, -0.6)$);

\node[below of=input, node distance=0.5cm] (trm) {(a) Transformer Layer.};

\end{tikzpicture}}
    \caption{Brief architecture of Transformer Encoder Layer and SASRec model.}
    \label{fig:model}
\end{figure}
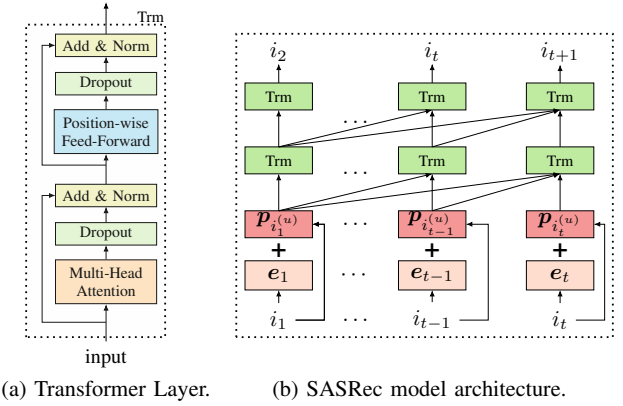
\textbf{SASRec} is different from FISM, since it utilizes sequential information when making recommendations~\cite{kang2018self}.
Equipped with Transformer encoder, SASRec is a left-to-right unidirectional model which predicts next item sequentially.
Transformer encoder consists of three sub-layers, an \textit{Embedding} layer, a \textit{Multi-Head Self-Attention} sub-layer and a \textit{Position-wise Feed-Forward Network} sub-layer.
The architectures of Transformer Encoder Layer and SASRec model are illustrated in Figure~\ref{fig:model}.

\textit{Embedding Layer}.
SASRec leverages position embedding $\bm{E} \in \mathbb{R}^{k \times d}$ to represent position information and it is added to the input embedding as:
\begin{equation}
    [\hat{\bm{S}}_u]_i^{\top}=\bm{p}_{i_t^{(u)}}+\bm{e}_t,
\end{equation}
where $\hat{\bm{S}}_u$ is the output matrix of Embedding Layer.
Here, we adopt the learnable position embedding for better performance.
However, it also induces a restriction on the maximum sequence length $L$.
To alleviate the problem, if $|\mathcal{S}_u| > L$, we truncate $\mathcal{S}_u$ to the last $L$ items:
\begin{equation}
    \mathcal{S}_u=\Bigl[i_{|\mathcal{S}_u|-L + 1}^{(u)},i_{|\mathcal{S}_u|-L+2}^{(u)},\dots,i^{(u)}_{|\mathcal{S}_u|}\Bigr].
\end{equation}

\textit{Multi-Head Self-Attention}.
Multi-head Self-attention~\cite{vaswani2017attention} focuses on integrating information from different subspaces at each position.
It consists of scaled dot-product attention mechanism, self-attention mechanism, and multi-head mechanism.
The scaled dot-product attention is defined as:
\begin{equation}
    \mathrm{Attention}(\bm{Q}, \bm{K},\bm{V})
    = \mathrm{softmax}\Bigl(\frac{\bm{Q}\bm{K}^{\top}}{\sqrt{d}}\Bigr)\bm{V},
\end{equation}
where $\bm{Q}$, $\bm{K}$, and $\bm{V}$ represent the queries, the keys, and the values, respectively.
When it comes to self-attention method, it leverages the same object and converts it to the query, the key, and the value with different linear projections as:
\begin{equation}
    \bm{H}=\mathrm{SA}(\hat{\bm{S}}_u)
    = \mathrm{Attention}(\hat{\bm{S}}_u\bm{W}^Q, \hat{\bm{S}}_u\bm{W}^K, \hat{\bm{S}}_u\bm{W}^V),
\end{equation}
where $\bm{W}^Q, \bm{W}^K, \bm{W}^V \in \mathbb{R}^{d \times d}$ are learnable parameters.
On the basis of self-attention, multi-head self-attention first projects $\hat{\bm{S}}_u$ into $h$ subspaces with different linear projections, then applies self-attention in parallel, and finally produces the output by concatenating the intermediates and projecting it once more.

\textit{Position-wise Feed-Forward Network}.
Though Multi-head Self-attention can aggregate all previous items' information,
it is mainly based on linear projections. 
To endow the model with nonlinearity and interactions
between different dimensions, we apply a Position-wise Feed-Forward Network to each position of the above sub-layer's output with sharing parameters:
\begin{equation}
\begin{aligned}
    \mathrm{PFFN}(\bm{H})
    &{=} \bigl[\mathrm{FFN}(\bm{h}_1)^{\!\!\top}\!; \mathrm{FFN}(\bm{h}_2)^{\!\!\top}\!; \dots;
    \mathrm{FFN}(\bm{h}_{|\mathcal{S}_u|})^{\!\!\top}\bigr] \\
    \mathrm{FFN}(\bm{h})
    &{=} \mathrm{RELU}(\bm{h}\bm{W}_1 + \bm{b}_1)\bm{W}_2 + \bm{b}_2.
\end{aligned}
\end{equation}

\textit{Stacking More Blocks}.
It is commonly beneficial to learn more complex item transition patterns by stacking more blocks. 
However, it becomes more difficult to converge as it goes deeper. 
We employ several mechanisms to alleviate this problem~\cite{vaswani2017attention}, a residual connection, a layer normalization, and a dropout module as follows:
\begin{equation}
    \mathrm{LayerNorm}(\bm{x} + \mathrm{Dropout}(\mathrm{sublayer}(\bm{x}))),
\end{equation}
where $\mathrm{sublayer(\cdot)}$ is the above sub-layer operations, including multi-head self-attention and feed-forward network.

After several Transformer blocks, the output at each position has already extracted information from previous items.
Since we intend to predict the item which the user $u$ will interact with at the timestamp $|\bm{S}_u|+1$, we set the representation for user $u$ as:
\begin{equation}
\bm{m}_u = [\texttt{Trm}^l(\mathcal{S}_u)]^{\top}_{|\bm{S}_u|},
    \label{eq:sasrec}
\end{equation}
where \texttt{Trm} is the abbreviation of the Transformer encoder mentioned above, $l$ is the number of \texttt{Trm} layers, and $\bm{m}_u$ is the ($|\bm{S}_u|$)-th row of the output.

\subsubsection{Parameters Learning}

The above parameters need to be learned to fit the users' historical interactions.
To fit the users' behaviors, different types of models adopt different strategies to derive training instances.
For non-sequential models, they usually predict each item $i \in \mathcal{R}^+$ with other items in $\mathcal{R}^+$.
When it comes to sequential recommendation models, they usually train the model by predicting the next item for each position in the input sequence.
Specifically, the target of the input sequence $[v_1, \cdots ,v_{|\mathcal{S}_u|-1}]$ is $[v_2, \cdots, v_{|\mathcal{S}_u|}]$ which is a shifted version of the input.
However, in the implicit feedback setting, we only have examples of what users prefer, but no examples of which kind of items they dislike, thus making it difficult to train the models.
To tackle this problem, we follow the settings in~\cite{he2018nais,kang2018self} to train the model with negative sampling and view the task as a binary classification problem. 
In this way, we treat the observed user-item interactions as positive instances and sample negative instances from the remaining unobserved ones.

Then the learning object is as follows:
\begin{equation}
    \mathcal{L} {=} {-}\frac{1}{M}\!\bigl(\!\!\! \sum_{(u,i)\in \mathcal{R}^+}\!\!\!\! \log \sigma(\hat{r}_{ui}^{\mathrm{\mathrm{UI}}}){+}\!\!\!\! \sum_{(u,j)\in \mathrm{Neg}}\!\!\!\! \log(1{-}\sigma(\hat{r}_{uj}^{\mathrm{\mathrm{UI}}}))\bigr) {+} \lambda\|\bm{\Theta}\|^2,
\end{equation}
where $M$ denotes the number of total training instances, $\mathrm{Neg} \subset \mathcal{R}^-$ denotes the negative sample set, $\hat{r}_{ui}^{\mathrm{\mathrm{UI}}}$ denotes the the vector dot product of user $u$ and item $i$, and $\sigma$ is the sigmoid function that converts the output to a probability. 
The hyper-parameter $\lambda$ controls the strength of $\ell_2$ regularization to prevent overfitting.
To optimize the objective function, we adopt mini-batch Adam~\cite{DBLP:journals/corr/KingmaB14}, a variant of Stochastic Gradient Descent~(SGD). 

\subsubsection{Candidate generation}
After inducing user representations, we can provide a personalized top-$N$ candidate list from the UI aspect.
It contains two stages, calculating the preference on the whole item set for each user and then selecting the top-$N$ items that are most likely to be clicked/purchased.

UI methods compute the preference for a user $u$ on an unobserved item $i$ (denoted by $\hat{r}_{ui}$) with dot product as follows:
\begin{equation}
\begin{aligned}
    \hat{r}_{ui}^{\mathrm{UI}} &=\bm{m}_u^{\top} \bm{q}_i,
\end{aligned}
\end{equation}
where $\bm{m}_u$ and $\bm{q}_i$ are embedding for user $u$ and embedding for item $i$, respectively.
And $\hat{r}_{ui}^{\mathrm{UI}}$ can be regarded as the correlation between user $u$ and item $i$.
Note that in some work~\cite{Kabbur:2013:FFI:2487575.2487589}, there exist two kinds of item embedding, input item embedding $\bm{p}_i$ and output item embedding $\bm{q}_i$ to add the model capacity.
In our framework, to reduce the model size and alleviate overfitting, we utilize a homogeneous item embedding, where $\bm{q}_i$ is equivalent to $\bm{p}_i$ like SASRec~\cite{kang2018self}.

Then given a specific user $u$, to suggest a small set of relevant candidates from a tremendous pool of items, 
we calculate the preference $\hat{r}_{ui}^{\mathrm{UI}}$ for each item $i$ in the set $\mathcal{R}_u^-$. 
Then, we generate a candidate set by selecting the highest $N$ items from them.

\subsection{User-based Component}
\subsubsection{Candidate generation}

As elaborated above, we have already acquired meaningful user representations and derived a candidate set from the global aspect.
In the second step, we concentrate on the perspective of local neighborhood information and recommend the products which the user $u$'s  neighbors prefer.

In this way, user-based component first computes the similarities between each user as follows:
\begin{equation}
    \mathrm{sim}(u, v) = \cos(\bm{m}_u, \bm{m}_v).
\end{equation}
Then it can detect a few similar neighbors for each user to construct her/his neighborhood.
We denote the neighborhood as $\mathcal{N}_u = \{v_1, v_2, \dots, v_{\beta}\}$, where $\beta$ is a hyper-parameter.
Note that $u\notin \mathcal{N}_u$ and neighbors in $\mathcal{N}_u$ are ordered descending by similarity scores.

With the information of neighborhood $\mathcal{N}_u$, we can derive another personalized top-N candidate list from the local aspect.
We set the user similarity $\mathrm{sim}(u, v)$ as the weight for the items user $v$ clicks/purchases, calculate another preference score on the whole item set, and then select the top-N items as candidates:
\begin{equation}
\label{user-score}
    \hat{r}_{ui}^{\mathrm{UU}}=\sum_{v\in\mathcal{N}_u}\delta_{vi}\mathrm{sim}(u,v)
\end{equation}
\begin{displaymath}
    \delta_{vi}=\left \{
    \begin{array}{lr}
        1, & i \in \mathcal{R}_v^+,\\
        0, & i \notin \mathcal{R}_v^+.\\
    \end{array}
\right.
\end{displaymath}
where $\hat{r}_{ui}^{\mathrm{UU}}$ denotes the preference of the user $u$ for item $i$ in the local perspective.
Furthermore, we assume that user $u$ will not click items in $\mathcal{R}_{u}^{+}$ once more, so we do not recommend items in $\mathcal{R}_{u}^{+}$.

\subsubsection{Complexity Analysis}
Here, we discuss why previous user-based models are not affordable in real-time scenarios, but our SCCF user-based component can adapt to it.

\textbf{Neighborhood user-based models}. The core of neighborhood user-based models is to estimate a user-user similarity matrix which is measured 
as follows:
\begin{equation}
    \mathrm{sim}(u,v)=\frac{|\mathcal{R}_u^+ \cap \mathcal{R}_v^+|}{|\mathcal{R}_u^+||\mathcal{R}_v^+|}
\end{equation}
Then they follow the equation \ref{user-score} to obtain the prediction. 
Calculating the similarity matrix is usually time consuming since it needs to compute similarities of high dimensional sparse vectors.
And its complexity is growing linearly with the number of items, which is not scalable.
Meanwhile, the neighborhood user-based models are transductive models.
If a user interacts with new items, these models need to recompute the similarity matrix,
whose high latency is not tolerated in real-world services.

\textbf{Factorized user-based methods}.
Factorized user-based methods factorize the user-item matrix to acquire user representations. 
Then they obtain the users' preference similar to our SCCF user-based component.
However, the factorized methods are transductive models as well.
If a user clicks/purchases a new item, these methods need to re-factorize the new user-item matrix to adapt to the latest user interest, which is infeasible in real-world services.

Our SCCF is based on an inductive UI model which can derive user representations from new behaviors by inference, not training.
Therefore, when a user clicks a new item, the user embedding can be inferred  without retraining in real-time, thus satisfying the need for online recommendation tasks. 

\subsection{The Integrating Component}
In this subsection, we discuss how to integrate the two candidate sets
and select the final top-$N$ candidates.
The two candidate lists for user $u$ are as follows:
\begin{equation}
    \begin{aligned}
    \mathcal{C}_{\mathrm{UI}}^u&=\{c_1^{\mathrm{UI}}, c_2^{\mathrm{UI}}, \dots, c_N^{\mathrm{UI}}\}\\
    \mathcal{C}_{\mathrm{UU}}^u&=\{c_1^{\mathrm{UU}}, c_2^{\mathrm{UU}}, \dots, c_N^{\mathrm{UU}}\},
    \end{aligned}
\end{equation}
which are ranked from the highest to the lowest by $\hat{r}_{ui}^{\mathrm{\mathrm{UI}}}$ and $\hat{r}_{ui}^{\mathrm{\mathrm{UU}}}$, respectively. %
Then, given a user $u$, it predicts the final score for all items in the two candidate sets as follows:
\begin{equation}
    \hat{r}_{ui}^{\mathrm{\mathrm{fi}}} = f(\bm{m}_u, \bm{q}_i, \hat{r}_{ui}^{\mathrm{\mathrm{UI}}}, \hat{r}_{ui}^{\mathrm{\mathrm{UU}}}),
\end{equation}
where the final score $\hat{r}_{ui}^{\mathrm{\mathrm{fi}}}$ takes both global and local information into account. 
We apply a multi-layer fully connected neural network to fuse the above features.
The input of it is the concatenation of four features:
\begin{equation}
\begin{aligned}
    \bm{input}_{ui} &= [\bm{m}_u \oplus \bm{q}_i \oplus \hat{r}_{ui}^{\mathrm{\mathrm{UI}}} \oplus \hat{r}_{ui}^{\mathrm{\mathrm{UU}}}]\\
    \tilde{r}_{ui}^{\mathrm{\mathrm{UI}}} &= (\hat{r}_{ui}^{\mathrm{\mathrm{UI}}} - \overline{\hat{r}_{ui}^{\mathrm{\mathrm{UI}}}})/\sigma{(\hat{r}_{ui}^{\mathrm{\mathrm{UI}}})} \\
    \tilde{r}_{ui}^{\mathrm{\mathrm{UU}}} &= (\hat{r}_{ui}^{\mathrm{\mathrm{UU}}} - \overline{\hat{r}_{ui}^{\mathrm{\mathrm{UU}}}})/\sigma{(\hat{r}_{ui}^{\mathrm{\mathrm{UU}}})},
\end{aligned}
\end{equation}
where $\oplus$ is the concatenation operation, $\hat{r}_{ui}^{\mathrm{\mathrm{UI}}}$ and $\hat{r}_{ui}^{\mathrm{\mathrm{UU}}}$ are all normalized preference scores, and the mean and standard deviation are calculated for each user $u$.
Then we stack several fully connected layers to the input embedding.

We optimize the network parameters on the following objective:
\begin{equation}
    \mathcal{L}_{\mathrm{I}}= {-}\!\sum_{u\in\mathcal{U}}\!\!
    \frac{1}{|\mathcal{C}_{\mathrm{I}}^u|}
    \Bigl(\sum_{i\in\mathcal{I}_u^+}\! \log\sigma(\hat{r}_{ui}^{\mathrm{fi}})
    {+}\!\!\sum_{i\in\mathcal{I}_u^-}\log(1{-}\sigma(\hat{r}_{ui}^{\mathit{\mathrm{fi}}})\!)\!\Bigr) {+}\lambda\|\bm{\Theta}\|^2
\end{equation}
where the set $\mathcal{C}^u_I = \mathcal{C}^u_{\mathrm{UI}}\cup\mathcal{C}^u_{\mathrm{UU}}$ denotes all the items in the two candidate sets, the item $i_u^+$ denotes the exact item which user $u$ clicks at the time $|\mathcal{S}_u| + 1$,
$\mathcal{I}_u^+=\{i_u^+\}\cap \mathcal{C}_I^u$ denotes the positive instance set,
and $\mathcal{I}_u^-=\mathcal{C}_I^u - \mathcal{I}_u^+$
denotes the negative instance set. 
If $i_u^+ \notin \mathcal{C}_I^u$, we will not calculate its two preference scores.
Therefore, we do not use this instance to train our integrating model.
This design exploits the four features in a fine grained, thus achieving well performance.

\section{Experiments}
In this section, we conduct plenty of experiments to answer the following research questions:
\begin{itemize}
    \item \textbf{RQ1.} How does our proposed SCCF framework perform compared to the state-of-the-art baselines?
    \item \textbf{RQ2.} Do the gains rely on the difference between UI methods results and user-based components results?
    \item \textbf{RQ3.} How does our user-based component in SCCF perform in the real-time settings?
    \item \textbf{RQ4.} How do the key hyper-parameters impact SCCF performance?
    \item \textbf{RQ5.} How does our SCCF perform on the e-commercial platform?
\end{itemize}

\subsection{Experiment Settings}
\subsubsection{Datasets}
We evaluate our methods on four publicly datasets.
\begin{itemize}[leftmargin=2em]
\setlength\itemsep{0.3em}
    \item \textbf{MovieLens}. 
    The first two datasets are both MovieLens \cite{Harper:2015:MDH:2866565.2827872} datasets, which are widely used for evaluating recommendation algorithms. 
    We use the version MovieLens-1M
    \footnote{\url{https://grouplens.org/datasets/movielens/1m/}}
    (\textbf{ML-1M}) and MovieLens-20M
    \footnote{\url{https://grouplens.org/datasets/movielens/20m/}}
    (\textbf{ML-20M}) that include 1 million and 20 million user ratings, respectively. 
    
    \item \textbf{Amazon}\footnote{\url{http://jmcauley.ucsd.edu/data/amazon/}}.
    The other two datasets are from amazon. 
    They are the series of datasets introduced in \cite{mcauley2015image}, which are split by top-level product categories in amazon and are notable for their high sparsity and variability. 
    In this work, we follow the settings in ~\cite{kang2018self} and adopt two categories, ``Beauty'' and ``Games''.
\end{itemize}
For dataset preprocessing, we follow the common practice in~ \cite{he2017translation,kang2018self,Sedhain:2016:PLM:3061053.3061158}.
We convert all numeric ratings or presence of a review to ``1'' and others to ``0''. 
It is worth mentioning that we follow the preprocessing procedure in \cite{he2017translation, kang2018self} for amazon datasets, which discards users and items with fewer than 5 related actions. 
And then to guarantee each user with enough interactions, we discard users with fewer than 5 actions once more. 
The processed data statistics are summarized in Table~\ref{tab:dataset}.

\begin{table}
\centering
  \caption{Dataset statistics (after preprocessing)}
  \label{tab:dataset}
  \begin{adjustbox}{max width=1.0\textwidth}
  \begin{tabular}{cccccc}
    \toprule
    Dataset &\#users &\#items &\#actions &avg.length & density\\
    \midrule
    ML-1M & 6040& 3416 &1.0M & 163.5 & 4.79\%\\
    ML-20M & 138493& 26744 &20M & 144.4 & 0.54\%\\
    Games & 29341& 23464&0.3M & 9.1 & 0.04\%\\
    Beauty & 40226 & 54542&0.4M & 8.8 & 0.02\%\\
  \bottomrule
\end{tabular}
\end{adjustbox}
\end{table}

\subsubsection{Evaluation}
We adopt the leave-one-out strategy to simulate the real-world applications and test our performance like ~\cite{he2018nais,he2017neural,kang2018self}.
For each user, we hold out the latest interaction as the test data, treat the item just before the last as the validation set and utilize others for training. 
Since we handle candidate generation tasks, we evaluate our model on the whole item set, sort the scores, and obtain the top $N$ highest items as the final results.
The performance is judged by Hit Ratio (HR)~\cite{Deshpande:2004:ITN:963770.963776} and Normalized Discounted Cumulative Gain (NDCG)~\cite{ricci2011introduction}.

$\bm{\mathrm{HR}@k}$ measures the proportion of cases that the desired item is among the top-$k$ items in test cases.
It is computed as: 
\begin{equation*}
    \mathrm{HR}@k = \frac{1}{\vert \mathcal{U} \vert}\sum_{u \in \mathcal{U}} \mathbbm{1}(\mathrm{rank}_{u, g_u} \leq k)
\end{equation*}
where $g_u$ is the ground truth item for user $u$, $\mathrm{rank}_{u, g_u}$ is the rank generated by the model for item $g_u$ and user $u$, and $\mathbbm{1}$ is an indicator function.
$\mathrm{HR}@k$ does not consider the actual rank of $g_u$ as long as it is among the top-$k$.

$\bm{\mathrm{NDCG}@k}$ is a position-aware metric which assigns larger weights on higher ranks. For next item recommendation, it is calculated as :
\begin{equation*}
    \mathrm{NDCG}@k = \frac{1}{\vert \mathcal{U} \vert}\sum_{u \in \mathcal{U}} \frac{2^{\mathbbm{1}(\mathrm{rank}_{u, g_u} \leq k)} - 1}{\log_2 (\mathrm{rank}_{u, g_u} + 1)}
\end{equation*}
In this work, we report HR and NDCG with $k=20,50,100$.

\subsubsection{Baselines}
To verify the effectiveness of SCCF, we validate it based on a diverse range of inductive UI models.
We select FISM~\cite{Kabbur:2013:FFI:2487575.2487589} and SASRec~\cite{kang2018self} as our UI component.

\begin{itemize}[leftmargin=2em]
\setlength\itemsep{0.3em}
    \item \textbf{FISM}. This is one of widely used shallow latent factor models. We use negative sampling to optimize the model~\cite{he2018nais}. 
    \item \textbf{SASRec}. This method is one of the state-of-the-art deep sequential models. It exploits the sequential information and utilizes self-attention modules to capture users' dynamic interests.
\end{itemize}

We also compare it with the following baselines:
\begin{itemize}[leftmargin=2em]
\setlength\itemsep{0.3em}
    \item \textbf{Pop}. This is a non-personalized benchmark.
It ranks items by their popularity---the number of interactions.
    \item \textbf{ItemKNN}~\cite{Sarwar:2001:ICF:371920.372071} and \textbf{UserKNN}~\cite{sarwar2000application}. They are traditional item-based CF and user-based CF methods proposed by \citeauthor{Sarwar:2001:ICF:371920.372071}. 
We use cosine similarity to derive the item or user similarity matrix. 
    \item \textbf{BPR-MF}~\cite{Rendle:2009:BBP:1795114.1795167}. BPR-MF optimizes matrix factorization with the pairwise Bayesian Personalized Ranking loss.
\end{itemize}

\subsubsection{Implementation Details and Parameter Settings}
We implement BPR-MF, FISM, SASRec, and our SCCF with \texttt{TensorFlow}\footnote{\url{https://www.tensorflow.org}}. 
We follow~\citet{he2018nais} to train FISM, which forms a mini-batch from all interactions of a randomly sampled user, rather than randomly sampling a fixed number of interactions.
Meanwhile, we train FISM without regularization and if overfitting is observed, we cope with it with early stop strategy like~\cite{he2018nais}.
However, when training SASRec model, we follow the same settings as ~\citet{kang2018self}, which uses dropout mechanism to avoid overfitting. 

To train the integrating model, we utilize each user's item just before the last as the training label.
Then, we randomly split ten percent of the whole users as the validation set to tune the integrating model.
We apply the early stop strategy to avoid overfitting.
When measuring the performance on the test set, we add all validation items and users back to the training set, and derive the final model parameters.

All parameters are initialized by truncated normal distribution in the range $[-0.01, 0.01]$.
We optimize the model by Adam optimizer with the learning rate of $0.001$, $\beta_1 = 0.9$, $\beta_2 = 0.999$, and linear decay of the learning rate.
For the embedding sizes, %
we test the values of $[16, 32, 64, 128]$.
Meanwhile, we test the effect of the number of user neighbors $\beta$ of $[ 50, 100, 200]$.
We set $\alpha=0.5$ for FISM model.
And the SASRec parameter settings are the same as \citeauthor{kang2018self}~\cite{kang2018self} with the layer number of 2 and head number of 1.
For fair comparison, we use the same maximum sequence length as in~\cite{kang2018self}, for ML-1M and ML-20M $N=200$, and for Games and Beauty $N=50$.
Since users' interests are dynamically changed, for FISM method, we leverage the recent 15 items to infer user embeddings.
And then for both methods, we recommend each user's latest 15 items to her/his similar users for user-based component. 

\subsection{Effectiveness of SCCF Framework(RQ1)}
We now compare the performance of our SCCF with the above baselines.
Table \ref{tab:result} summarizes the best results of all models on four datasets.
Note that the improvement columns are the performance of SCCF relative to its base UI components.
It can be observed that:
\begin{itemize}[leftmargin=2em]
\setlength\itemsep{0.3em}
\item The non-personalized methods, Pop and ItemKNN methods, exhibit the worst performance since they ignore users' unique interests.
FISM and BPR-MF outperform the above methods since they provide personalized recommendations.
Among all the baselines, SASRec achieves the state-of-the-art performance on all datasets, since it leverages sequence information and utilizes powerful self-attention mechanism. 
 
\item Despite direct training is conducted on the UI models, user-based component is capable of achieving comparable performance.
When it comes to FISM baseline, the metric of user-based component is even higher than original FISM on all datasets.
In contrast, for SASRec baseline, it outperforms SASRec$_\mathrm{UU}$ on all the datasets.
It may be the fact that SASRec captures more complex item semantic relations than FISM.

\item According to the results, it is obvious that our proposed framework SCCF outperforms all the UI models on all datasets in terms of all evaluation metrics.
SASRec$_\mathrm{SCCF}$ gains 8.90\% HR@20, 1.94\% HR@100, 12.75\% NDCG@20, and 7.42\% on NDCG@100 on average against the start-of-the-art baseline SASRec. 
Our SCCF framework also obtains dramatically remarkable improvements over the FISM model on all datasets.
The improvements are even 16.14\% on HR@100 and 30.84\% on NDCG@100 on average. 
These experiments verify
the effectiveness of our user-based component in SCCF framework.

\item We also observe one phenomenon exhibiting in the experiment results.
Our SCCF framework promotes SASRec most remarkable on ML-20M dataset than other datasets.
SASRec needs to learn long sequence dependencies with enough training instances. 
Therefore, with less training instances (ML-1M) and fewer interactions per user (amazon datasets), it is difficult for SASRec to derive meaningful representations, thus difficult to identify accurate user neighborhood.

\end{itemize}
\begin{table*}[]
\setlength{\tabcolsep}{0.65em}
\renewcommand{\arraystretch}{1.3}
    \centering
    \caption{Performance comparison of different methods on top-$N$ recommendation. Bold scores are the best in method group. Improvements are calculated over UI methods.} %
    \begin{adjustbox}{max width=1.0\textwidth}
    \begin{threeparttable}
        \begin{tabular}{l| l| c c c c|c c c c| c c c c }
        \toprule
        Datasets & Metric & Pop & ItemKNN & UserKNN & BPR-MF & FISM & FISM$_{\mathit{UU}}$ & FISM$_{\mathit{SCCF}}$ & Improv. & SASRec & SASRec$_{\mathit{UU}}$ & SASRec$_{\mathit{SCCF}}$ & Improv.\\
        \midrule
        \multirow{6}{*}{ML-1M} 
        & HR@20 & 0.0596 & 0.1179 & 0.1308&0.1424 & 0.2189 & 0.2374 & 0.2677 & 22.29\% & 0.3447 & 0.2639 & \textbf{0.3613} & ~~4.82\% \\
        
         & HR@50 & 0.1356 & 0.2265 & 0.2444&0.2659 & 0.3740 & 0.3891 & 0.4041 & ~~8.05\% & 0.5227 & 0.4346 & \textbf{0.5281} & ~~1.03\%\\
         
         & HR@100 & 0.2190 & 0.3419 &0.3593 &0.4060 & 0.5017 & 0.5066 & 0.5315 & ~~5.94\% & 0.6409 & 0.5710 & \textbf{0.6411} & ~~0.03\%\\
         
         & NDCG@20 & 0.0224 & 0.0471 &0.0513 &0.0561 & 0.0891 & 0.0998 & 0.1177 & 32.10\% & 0.1521 & 0.1122 & \textbf{0.1654} & ~~8.74\%\\
         
         & NDCG@50 & 0.0372 & 0.0684 &0.0737 &0.0804 & 0.1200 & 0.1298 & 0.1447 & 20.58\% & 0.1874 & 0.1460 & \textbf{0.1985} & ~~5.92\%\\
         
         & NDCG@100 & 0.0507 & 0.0870 &0.0923 &0.1030 & 0.1408 & 0.1489 & 0.1653 & 17.40\% & 0.2067 & 0.1682 & \textbf{0.2169} & ~~4.93\%\\
         \midrule
        \multirow{6}{*}{ML-20M} 
        & HR@20 & 0.0675 & 0.1279 & -\tnote{1}
        &0.1099 & 0.2022 & 0.2220 & 0.2558 & 26.51\% & 0.2524 & 0.2327 & \textbf{0.2939} & 16.44\% \\
        
         & HR@50 & 0.1154 & 0.2302 & -&0.2044 & 0.3325 & 0.3558 & 0.3887 & 16.90\% & 0.4075 & 0.3763 & \textbf{0.4388} & ~~7.68\%\\
         
         & HR@100 & 0.1832 & 0.3524 & -&0.3061 & 0.4502 & 0.4641 & 0.4985 & 10.73\% & 0.5415 & 0.4967 & \textbf{0.5569} & ~~2.84\%\\
         
         & NDCG@20 & 0.0274 & 0.0520 &-& 0.0449 & 0.0852 & 0.0938 & 0.1183 & 38.85\% & 0.1065 & 0.0979 & \textbf{0.1346} & 26.38\%\\
         
         & NDCG@50 & 0.0370 & 0.0721 &- &0.0635 & 0.1109 & 0.1203 & 0.1446 & 30.39\% & 0.1372 & 0.1263 & \textbf{0.1634} & 19.10\%\\
         
         & NDCG@100 & 0.0479 & 0.0919 &- &0.0800 & 0.1300 & 0.1379 & 0.1624 & 24.92\% & 0.1589 & 0.1459 & \textbf{0.1825} & 14.85\%\\
         \midrule
        \multirow{6}{*}{Games} & 
        HR@20 & 0.0333 & 0.0736 &0.1227 &0.1210 & 0.0974 & 0.1157 & 0.1413 & 45.07\% & 0.1325 & 0.1241 & \textbf{0.1464} & 10.49\%\\
         
         & HR@50 & 0.0591 & 0.1189 &0.1852 &0.2048 & 0.1684 & 0.1860 & 0.2215 & 31.53\% & 0.2300 & 0.2035 & \textbf{0.2468} & ~~7.30\%\\
         
         & HR@100 & 0.0919 & 0.1637 &0.2338 &0.2888 & 0.2445 & 0.2469 & 0.2879 & 17.75\% & 0.3260 & 0.2755 & \textbf{0.3405} & ~~4.45\% \\
         
         & NDCG@20 & 0.0155 & 0.0315 & 0.0546&0.0500 & 0.0399 & 0.0483 & \textbf{0.0639} & 60.15\% & 0.0536 & 0.0525 & 0.0599& 11.75\% \\
         & NDCG@50 & 0.0206 & 0.0405 & 0.0670&0.0665 & 0.0540 & 0.0622 & \textbf{0.0798} & 47.78\% & 0.0728 & 0.0682 & \textbf{0.0798} & ~~9.62\%\\
         & NDCG@100 & 0.0259 & 0.0477 &0.0749 &0.0801 & 0.0663 & 0.0721 & 0.0906 & 36.65\% & 0.0883 & 0.0798 & \textbf{0.0950} & ~~7.59\%\\
         \midrule
        \multirow{6}{*}{Beauty} 
        & HR@20 & 0.0134 & 0.0422 & 0.0676&0.0600 & 0.0425 & 0.0574 & 0.0628 & 47.76\% & 0.0648 & 0.0505 & \textbf{0.0673} &3.86\%\\
         
         & HR@50 & 0.0248 & 0.0621 & 0.0942&0.1010 & 0.0705 & 0.0882 & 0.0983 & 39.43\% & 0.1045 & 0.0855 & \textbf{0.1072} & 2.58\%\\
         
         & HR@100 & 0.0398 & 0.0829 &0.1160 &0.1390 & 0.1006 & 0.1107 & 0.1309 & 30.12\% & 0.1460 & 0.1165 &  \textbf{0.1466} &~~0.42\%\\
         
         & NDCG@20 & 0.0050 & 0.0198 &0.0320 &0.0250 & 0.0173 & 0.0248 & 0.0278 & 52.63\% & 0.0289 & 0.0215 & \textbf{0.0301} & ~~4.15\%\\
         
         & NDCG@50 & 0.0072 & 0.0237 & 0.0373&0.0332 & 0.0228 & 0.0310 & 0.0348 & 52.63\% & 0.0367 & 	0.0284 & \textbf{0.0380} & ~~3.54\%\\
         
         & NDCG@100 & 0.0097 & 0.0271 &0.0408 &0.0393 & 0.0277 & 0.0346 & 0.0400 & 44.40\% & 0.0434 & 0.0334 & \textbf{0.0444} & ~~2.30\%\\
        \bottomrule
        \end{tabular}
        \begin{tablenotes}
\item[1] Running UserKNN on ML-20M will lead to timeout.
\end{tablenotes}
    \end{threeparttable}
    \end{adjustbox}
    \label{tab:result}
\end{table*}

\subsection{Complementary issues(RQ2)}
To explore whether UI and user-based components recommend items in different ways, we investigate their recommendation results.
We make this analysis on the experiment of SASRec$_\mathrm{SCCF}$ on the ML-20M dataset.
For ground truth, we compute the cosine similarity between the user embedding and the target item embedding for each user.
When it comes to the UI and user-based components, we calculate the cosine scores between the user and each item in the candidate sets.
Then for each user, we average these values to obtain the score of the whole candidate set.
Finally, we summarize the number of users of each value.

\begin{figure}
  \centering
  \includegraphics[width=0.9\linewidth]{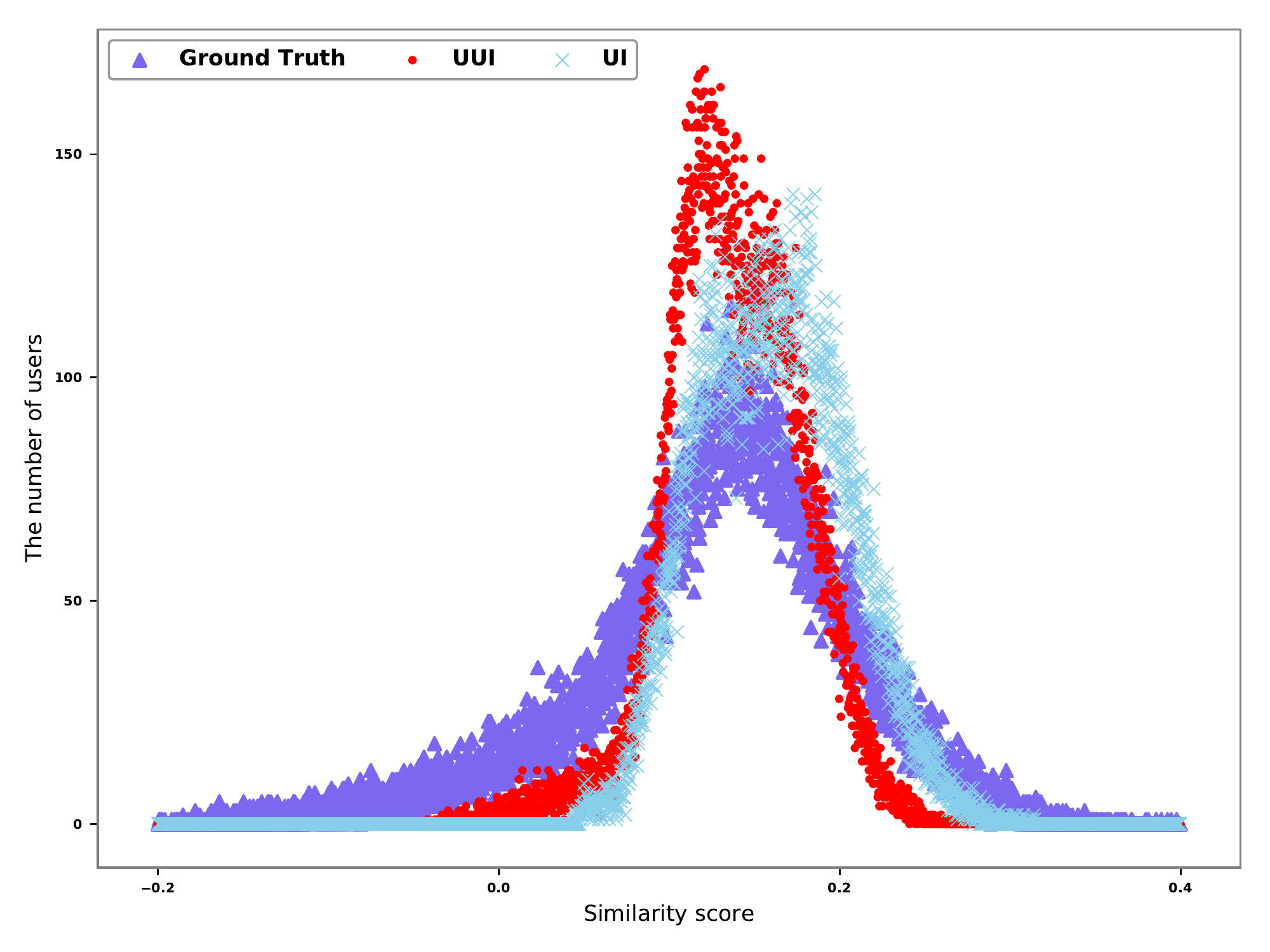}
  \caption{The number of users with similarity score with SASRec methods on ML-20M dataset. 
  Each point is the similarity score between the user and the item.
  Note that the UI and user-based (UUI) scores are calculated by averaging the user-item similarity scores of all the candidate items.}
  \label{Diff}
\end{figure}

The results are visualizing in Figure~\ref{Diff}.
We observe that the mean score of UI component similarity distribution is a litter higher than ground truth, while user-based component is lower than the target item distribution.
This phenomenon reveals that there really exists significantly difference between UI methods and user-based component.
UI method usually tends to provide items with higher similarity, since it is optimized by minimizing the distance between user representations and interacted item representations.
In this way, they fail to recommend items which may be a little far from the specific user.
In contrast, user-based component can capture more local information in user neighborhood and recommend items which are not very similar to the users.
Therefore, UI and user-based components can complement each other in global and local manners, which enhances our SCCF to achieve better performance.

\subsection{Efficiency Analysis(RQ3)}
In this section, we conduct experiments to verify the efficiency of our SCCF user-based component in the real-time settings.
Since user-based baseline models cannot be applied in large-scale applications, we simulate the real-time settings on public datasets. 
For the fact that factorized user-based methods are transductive models, they need to re-factorized the user/item matrix and cost more time than other methods.
Therefore, we just compare with UserKNN and explore the compromise between inferring new representations and identifying neighbors.
We report the time consumed by each method to make new predictions when a user interact with a new item.
Results are the average of all users.
Experiments are conducted on a V100 GPU with 16GB memory.
To make fair comparison, we just utilize ML-1M and Video datasets since baseline methods require more GPUs with extra communication on other larger datasets.
We apply SCCF on SASRec for comparison.

The results are displayed in Table~\ref{tab:time}.
We observe that SCCF user-based component is more efficient and scalable than UserKNN.
Though our SCCF needs to spend extra time to infer user representations dynamically, it identifies user neighborhood more quickly and scalable with low dimentional vectors.
And the latency of UserKNN is growing at least linearly with the number of items, thus restrcting it to be applied in real world applications with millions even billions of items.

\input{groupplot.tex}

\begin{table}
    \centering
    \caption{Performance comparison SCCF user-based components and other user-based baselines in real-time settings. 
    }
    \begin{adjustbox}{max width=\linewidth}
        \begin{tabular}{c l c c c c}
        \toprule
        \multirow{2}{*}{Datasets}
        & \multirow{2}{*}{}
        & \multicolumn{2}{c}{Methods}\\ \cmidrule(lr){3-4}
        &&UserKNN & SCCF \\
        \midrule
        \multirow{3}{*}{ML-1M} 
        & Inferring time (ms) & 0 & 1.66 \\
        
        & Identifying time (ms)& 6.83 & 0.72\\
         
        & Total time (ms) & 6.83 & 2.38\\
         
         \midrule
        \multirow{3}{*}{Videos} 
        & Inferring time (ms)& 0 & 0.79\\
        
        & Identifying time (ms)& 51.95 & 0.75\\
         
        & Total time (ms)& 51.95 & 1.54\\
         
        \bottomrule
        \end{tabular}
    \end{adjustbox}
    \label{tab:time}
\end{table}

\subsection{Hyper-parameter Study(RQ4)}
In this section, we analyze how the hyper-parameters impact the performance, including the hidden dimension %
and the size of user neighborhood $N_u$.
We examine the influence of one hyper-parameter each time by fixing the remaining hyper-parameters.
We report HR@50 and NDCG@50 as metrics for the hidden dimension experiments and we just report NDCG@50 for the neighborhood experiments due to the space limitation.

First, we examine the impact of hidden dimension. 
Figure~\ref{fig:dimension} demonstrates the performance of different embedding sizes %
while keeping other hyper-parameters unchanged.
We observe a few trends with the variation of embedding sizes.
First, the performance of each SCCF instance tends to converge as the dimension increases on most datasets.
But a large dimension does not necessarily promise better performance due to the overfitting problem.
On some datasets, such as Games, high dimension even sacrifices the performance.
Second, the performance trend that our SCCF outperforms their base UI component is consistent with different embedding sizes. 
Even with a relatively small hidden size, our SCCF framework still outperforms its base component.

Then we study how the neighbor number $\beta$ influences the performance.
Table~\ref{tab:user} demonstrates the performance of different neighbor numbers. SCCF achieves stable improvements over different neighborhood sizes. 
We also observe that enlarging user neighborhood may decrease the model quality, since it may introduce noise users.

\begin{table}
    \centering
    \caption{Performance comparison of differen neighbor numbers $\beta$ on NDCG@50. Bold scores are the best in groups. 
    } %
    \begin{adjustbox}{max width=\linewidth}
        \begin{tabular}{c c c c c c}
        \toprule
        \multirow{2}{*}{Methods}
        & \multirow{2}{*}{Neighbors}
        & \multicolumn{4}{c}{Datasets}\\ \cmidrule(lr){3-6}
        &&ML-1M & ML-20M & Games & Beauty \\
        \midrule
        \multirow{3}{*}{FISM$_\mathrm{UI}$} 
        & $\beta=50$ & 0.1200 & 0.1109 & 0.0540 & 0.0228 \\
        
        & $\beta=100$ & 0.1200 & 0.1109 & 0.0540 & 0.0228\\
         
        & $\beta=200$ & 0.1200 & 0.1109 & 0.0540 & 0.0228\\
         
         \midrule
        \multirow{3}{*}{FISM$_\mathrm{UU}$} 
        & $\beta=50$ & 0.1289 & 0.1179 & 0.0587 & 0.0289 \\
        
        & $\beta=100$ & 0.1298 & 0.1203 & 0.0622 & 0.0310\\
         
        & $\beta=200$ & 0.1287 & 0.1205 & 0.0631 & 0.0308\\
         
         \midrule
        \multirow{3}{*}{FISM$_\mathrm{SCCF}$} 
        & $\beta=50$ & 0.1433 & 0.1436 & 0.0778 & 0.0353 \\
        
        & $\beta=100$ & \textbf{0.1447} & \textbf{0.1446} & \textbf{0.0798} & 0.0348\\
         
        & $\beta=200$ & \textbf{0.1447} & 0.1428 & 0.0788 & \textbf{0.0349}\\
        \midrule
        \midrule
        \multirow{3}{*}{SASRec$_\mathrm{UI}$} 
        & $\beta=50$ & 0.1874 & 0.1372 & 0.0728 & 0.0367 \\
        
        & $\beta=100$ & 0.1874 & 0.1372 & 0.0728 & 0.0367\\
         
        & $\beta=200$ & 0.1874 & 0.1372 & 0.0728 & 0.0367\\
         
         \midrule
        \multirow{3}{*}{SASRec$_\mathrm{UU}$} 
        & $\beta=50$ & 0.1454 & 0.1245 & 0.0645 & 0.0312 \\
        
        & $\beta=100$ & 0.1460 & 0.1263 & 0.0682 & 0.0284\\
         
        & $\beta=200$ & 0.1444 & 0.1248 & 0.0691 & 0.0238\\
         
         \midrule
        \multirow{3}{*}{SASRec$_\mathrm{SCCF}$} 
        & $\beta=50$ & 0.1972 & \textbf{0.1651} & 0.0775 & \textbf{0.0383} \\
        
        & $\beta=100$ & \textbf{0.1985} & 0.1634 & \textbf{0.0798} & 0.0380\\
         
        & $\beta=200$ & 0.1973 & 0.1643 & 0.0783 & \textbf{0.0383}\\
        \bottomrule
        \end{tabular}
    \end{adjustbox}
    \label{tab:user}
\end{table}

\subsection{Online A/B Test Result(RQ5)}
Previous offline experiment results have demonstrated the superiority of our SCCF framework. 
In this subsection, we deploy it on the mobile Taobao, which is the largest e-commerce platform in China.  
The experiments are conducted at What You May Like, the major recommendation scenario in Taobao.

It is worth mentioning the two-step process has been widely applied in the large scale
recommendation scenarios~\cite{Davidson:2010:YVR:1864708.1864770,Covington:2016:DNN:2959100.2959190}. 
We also adopt this procedure on our platform.
To make a fair comparison, we keep all downstream modules unchanged except the candidate generation module.
We restrict the candidate set to 500 items and provide this set as input for the following stage.
The baseline we deployed online is a deep model similar to the method 
proposed by \citeauthor{Covington:2016:DNN:2959100.2959190}~\cite{Covington:2016:DNN:2959100.2959190}.
We plugin our SCCF and provide exactly the same number of candidates.
We randomly split some online users equally into two separate buckets. 
For users in the bucket A (i.e., the baseline group), we derive candidate generation step with the base approach.
While for users in the bucket B (i.e., the experimental group), the candidates are suggested by our SCCF.

The experiment is deployed online for one week from 8. Aug to 14. Aug in 2019.
Each bucket contains about one million user's daily traffic.
Different from offline metrics, the performance is judged by total user clicks and trades, which have strong correlations with the profit of corporations.
Table~\ref{tab:online_exp} shows that SCCF outperforms the baseline on both total clicks and total trade volume.
The number of clicks is improved by 2.5\% and trade volume is improved by 2.3\% on average. 
The improvements are remarkable in the industrial online environment. 

\begin{table}
\centering
  \caption{Online experiment results}
  \label{tab:online_exp}
  \begin{adjustbox}{max width=\linewidth}
  \begin{tabular}{ccc}
    \toprule
    Metric &\#Clicks &\#Trades \\
    \midrule
    Lift Rate & 2.5\%& 2.3\%\\
  \bottomrule
\end{tabular}
\end{adjustbox}
\end{table}

\section{Conclusion}
In this paper, we explore the effectiveness of the user-based methods and we propose an efficient framework SCCF for candidate generation.
SCCF combines information from both global and local perspectives captured by the UI and the user-based components, respectively.
Meanwhile, the user-based component of SCCF can be utilized in real time by dynamically inferring the user representations.
In this way, SCCF is the first to address this real-time problem of user-based methods.
Most importantly, it can be easily applied to existing inductive UI methods.
The proposed framework is verified on both publicly datasets and an online industrial environment. 
Extensive experiment results show that the proposed framework achieves significant improvements and outperforms the start-of-the-art baselines.

In the future, we will investigate how to incorporate side information such as user profile to identify similar users for each user.
Since user profile usually contains general user information, it can help us identify more accurate user preference.
In addition, we will also attempt to apply our approach to the ``ranking'' step besides ``candidate generation'' step,
since existing methods only consider user-item relation to predict the score for each candidate in the ``ranking'' step.

\section*{Acknowledgement}
This work is supported by the National Key Research and Development Program of China (No. 2018YFB1004403), the National Natural Science Foundation of China under Grant (No. 61832001), Beijing Academy of Artificial Intelligence (BAAI), and Alibaba-PKU joint program.

\bibliographystyle{IEEEtranN}
\bibliography{main}

\end{document}